\documentclass[10pt,conference]{IEEEtran}
\IEEEoverridecommandlockouts

\usepackage{cite}
\usepackage{amsmath,amssymb,amsfonts}
\usepackage{algorithmic}
\usepackage{graphicx}
\usepackage{textcomp}
\usepackage{xcolor}
\def\BibTeX{{\rm B\kern-.05em{\sc i\kern-.025em b}\kern-.08em
    T\kern-.1667em\lower.7ex\hbox{E}\kern-.125emX}}

\usepackage[frozencache,cachedir=minted-cache]{minted}
\usepackage{array}

\usepackage{enumitem}
\usepackage{multirow}
\usepackage{makecell}
\usepackage[textsize=small]{todonotes}
\usepackage{listings}
\usepackage[hidelinks]{hyperref}
\usepackage{booktabs}
\usepackage{multirow}
\usepackage{url}
\usepackage{xspace}

\newcommand{\researchquestion}[2]{%
  \vspace{4mm}
  \noindent
  \setlength{\fboxsep}{0pt}%
  \colorbox{white}{%
    \hspace{0pt}%
    \textcolor{gray!50}{\vrule width 1.5pt height \dimexpr\ht\strutbox+\dp\strutbox\relax}%
    \hspace{1em}%
    \begin{minipage}[t]{0.95\linewidth}%
      \textbf{#1}\ #2
    \end{minipage}%
  }%
  \par
  \vspace{4mm}%
}

\newcommand\RQAnswer[2]{%
  \vspace{\topsep}
  \noindent
  \setlength{\fboxsep}{4pt}%
  \colorbox{gray!10}{%
    \begin{minipage}[t]{0.95\linewidth}%
      \textbf{#1}\ #2
    \end{minipage}%
  }%
  \par
  \vspace{\topsep}%
}

\definecolor{SnippetOriginal}{HTML}{4B5563}
\definecolor{SnippetCueVaried}{HTML}{2563EB}
\definecolor{SnippetImplementationVaried}{HTML}{EA580C}

\definecolor{ConsistentBoth}{HTML}{7C3AED}
\definecolor{ConsistentCue}{HTML}{2563EB}
\definecolor{ConsistentExecution}{HTML}{EA580C}
\definecolor{ConsistentNeither}{HTML}{9CA3AF}

\newcommand\alignedSn{aligned\xspace}
\newcommand\cuevaried{cue-varied\xspace}
\newcommand\implementationvaried{implementation-varied\xspace}

\newcommand\bothconsistent{both-consistent\xspace}
\newcommand\cueconsistent{cue-consistent\xspace}
\newcommand\executionconsistent{execution-consistent\xspace}
\newcommand\neither{neither\xspace}

\begin{document}

\title{A Mechanistic Lens on Semantic Conflicts: Using Activation Patching to Understand LLM Behavior}












\author{
\IEEEauthorblockN{Youssef Abdelsalam, Norman Peitek, Anna-Maria Maurer, Marvin Wyrich, and Sven Apel}
\IEEEauthorblockA{\textit{Saarland Informatics Campus, Saarland University}\\
Saarbr\"ucken, Germany
}
}

\maketitle

\begin{abstract}
Large language models (LLMs) are increasingly used in software-engineering tasks processing executable code and non-executable semantic cues such as comments or identifiers.
These two sources of information can conflict, leading to situations where the semantic cues suggest different program behavior than the code itself.
It remains unclear how such semantic conflicts affect LLM behavior and which source of information dominates their outputs.

We present the first controlled, mechanistic study of LLM behavior under semantic conflicts.
To this end, we construct 45 Python snippet triplets that isolate conflicts by varying either semantic cues or implementation while keeping token-aligned pairs for causal intervention.
We evaluate four open-weight LLMs on two tasks---final-output prediction and unit-test generation---using both behavioral performance measures and residual-stream activation patching to identify token-layer states that causally contribute to differences in LLM behavior between aligned and conflicting inputs.

Our results show that semantic conflicts significantly reduce execution-grounded correctness in both tasks and that all tested LLMs frequently follow (misleading) semantic cues.
Residual-stream activation patching reveals a consistent pattern for final-output prediction: The changed cue/code region and a small set of intermediate tokens carry most of the recoverable causal signal before being aggregated near the output readout.
For unit-test generation, this pattern extends beyond the prompt, showing that conflict-related information is not only recoverable at prompt sites but also at generated assertion sites before producing expected values.
Overall, our findings show that semantic conflicts affect both program comprehension and downstream tasks, with the relevant information concentrated in a small number of causally active residual-stream states, and demonstrate a framework for mechanistically analyzing how LLMs integrate different sources of code-related information under controlled semantic variations.
\end{abstract}


\begin{figure*}
    \centering
    \includegraphics[width=1\linewidth]{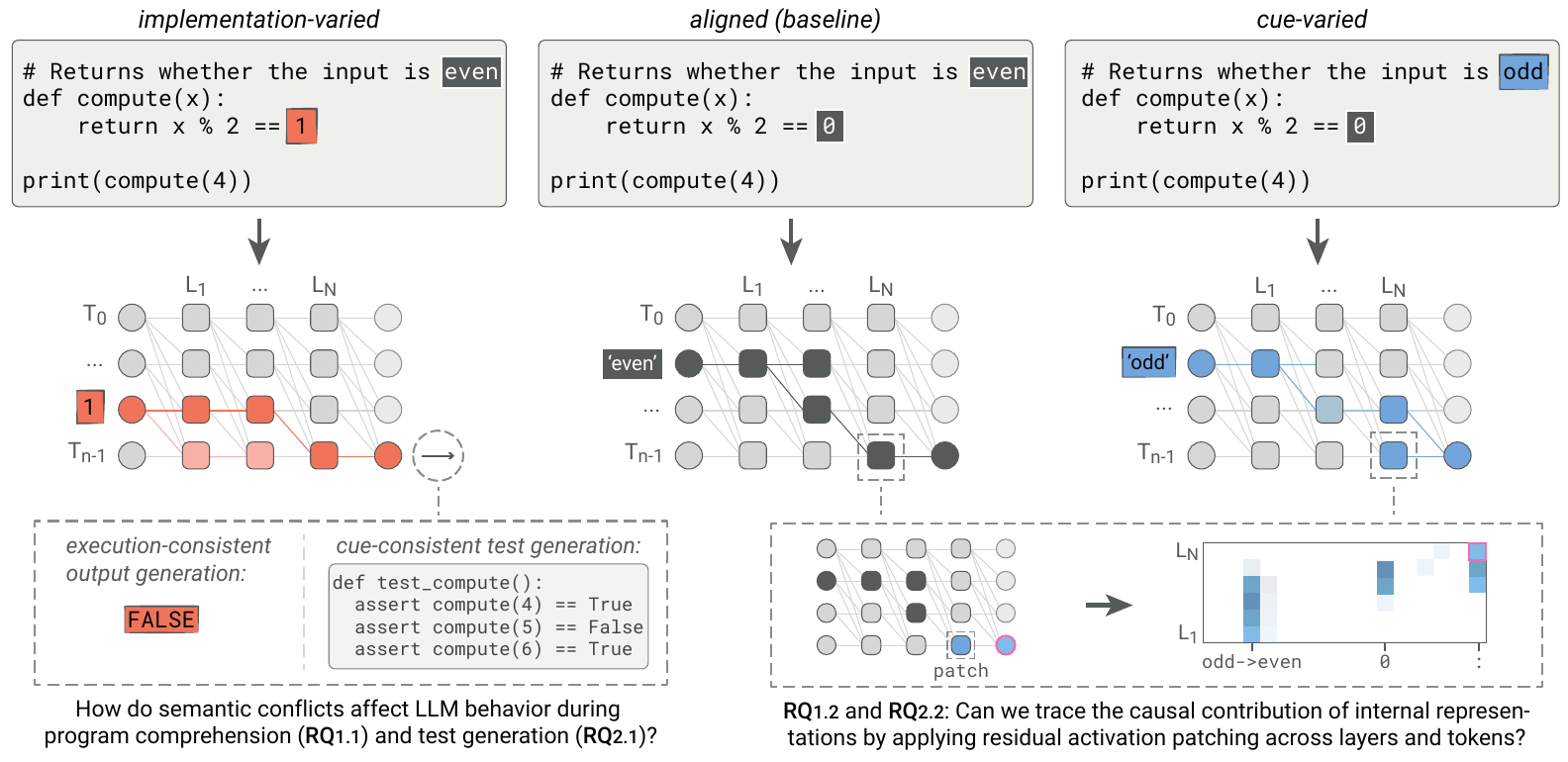}
    \caption{Overview of our experimental framework. We construct 45 Python snippet triplets (aligned, implementation-varied, and cue-varied) and provide them together with either an output prediction task (RQ\textsubscript{1}) or a unit-test generation task (RQ\textsubscript{2}) to four LLMs. We then (i) assess how semantic conflicts shift model behavior relative to execution and cues, and (ii) identify the token-layer sites that causally drive these shifts via residual activation patching.}
    \label{fig:ExperimentOverview}
\end{figure*}

\section{Introduction}
\label{sec:introduction}

Large language models (LLMs) are increasingly used in software-engineering workflows~\cite{Hou2024Large, Zhang2026Survey}, including test generation~\cite{Wang2024Testing, Chang2025Systematic}, program repair~\cite{Fan2023Repair, Yin2024ThinkRepair}, and agentic development~\cite{He2025MultiAgent, Liu2026Agents}.
These workflows expose LLMs to both executable code and non-executable semantic cues, such as comments, identifiers, and other natural-language texts.
While these sources typically agree in well-maintained code~\cite{Parnas2010, Huang2025}, they often diverge in practice due to outdated documentation, misleading identifiers, or evolving implementations~\cite{Aghajani2020, Stulova2020, Huang2025}, resulting in \emph{semantic conflicts}.
Such conflicts are inherently ambiguous, as neither source can be assumed authoritative without external validation.

This renders LLMs distinctive among software-engineering tools, as they jointly process executable implementations and natural-language cues.
In fact, LLMs often rely on semantic cues in their reasoning~\cite{Gao2023, Le2025}.
When conflicts arise, it remains unclear which source LLMs prioritize and how this affects downstream tasks such as test generation, program repair, and agentic development.
Prior work has documented many LLM failures in coding tasks~\cite{Tambon2025, Dou2026}, but behavioral outputs alone cannot explain how conflicting information is internally represented or how it influences downstream software-engineering tasks.

We address this gap by studying LLM behavior under semantic conflicts in two tasks: predicting program outputs and generating unit tests.
We introduce an experimental framework that combines behavioral evaluation with mechanistic interpretability.
Using residual-stream activation patching, an intervention-based method commonly used in mechanistic interpretability~\cite{zhang2024bestpractices, heimersheim2024patching}, we causally trace where conflict-relevant representations emerge within the model.
To this end, we construct paired inputs in which identical code is presented with either consistent (aligned) or inconsistent (conflicting) semantic cues, allowing us to patch residual states between these conditions and identify token-layer sites whose internal states are sufficient to shift the LLM's outputs.

Our dataset consists of 45 minimal, token-matched Python triplets: an aligned baseline and two conflicting variants created by modifying either the implementation or a semantic cue.
Across four open-weight autoregressive transformer models, we find that semantic conflicts substantially reduce execution-grounded correctness in both tasks.
Models frequently follow misleading semantic cues, producing incorrect outputs or tests.
Mechanistically, we observe a staged pattern: early layers are sensitive to modified cue/code regions, middle layers to a sparse set of intermediate tokens, and later layers to the final input context.

To our knowledge, this is among the first applications of causal mechanistic interpretability methods in a software-engineering context.
Beyond characterizing the effects of semantic conflicts, our framework shows how such methods can localize behaviorally relevant signals within the residual stream, narrowing the search space for future analyses (e.g., path patching~\cite{Goldowsky2023} or circuit tracing~\cite{Conmy2023}).
This opens avenues for diagnosing LLM behavior, improving reliability, and detecting conflicts between cue- and execution-consistent signals before they affect outputs during code-related tasks.

In summary, we make the following contributions:

\begin{itemize}[leftmargin=3ex]
    \item An experimental framework to study semantic conflicts between cues and implementations, introducing activation patching for software-engineering research.
    \item A behavioral analysis of 45 Python triplets showing that semantic conflicts significantly reduce correctness in both output prediction and generated unit tests.
    \item A causal mechanistic analysis identifying token-layer states that shift outputs between aligned and conflicting inputs. 
    \item A replication package, including analysis scripts and 45 token-aligned code snippet triplets, to support future research~\cite{zenodo:dataset}.
\end{itemize}

Figure~\ref{fig:ExperimentOverview} summarizes the resulting study design, connecting the aligned and conflicting snippet variants to the behavioral measurements and residual-stream interventions used throughout the paper.

\section{Background and Related Work}
\label{sec:background-related-work}

In this section, we provide the necessary background and summarize prior work related to our study. We review three areas: (1)~LLMs in software engineering, (2)~mechanistic interpretability of LLMs, and (3)~semantic conflicts between executable code and non-executable software artifacts.

\subsection{LLMs in Software Engineering}
\label{sec:rw-llms-se}

LLMs for code generation are commonly evaluated through external behavioral criteria, especially execution-based functional correctness. For example, \textsc{HumanEval} introduced a benchmark of Python programming problems evaluated by unit tests and popularized $\text{pass@k}$-style evaluation~\cite{chen2021evaluating}. The \textsc{MBPP} benchmark similarly evaluates synthesis of short Python programs from natural-language task descriptions and tests~\cite{austin2021program}. Subsequent work has shown that such test-based evaluation depends on the size and quality of the test set, showing that many LLM-generated solutions previously counted as correct fail under more rigorous testing~\cite{liu2023evalplus}. In any case, these benchmarks and extensions establish functional correctness as a central evaluation target for code LLMs, but which is often assessed only for the final code output.

LLMs are also increasingly applied for test generation. Recent benchmarks evaluate generated tests using a broad range of criteria, such as validity, coverage, mutation score, pass rate, and fault-detection ability~\cite{Wang2024Testing,Chang2025Systematic,Wang2025TestEval}. \textsc{ULT} is a benchmark for function-level unit-test generation from real-world Python functions. It reports accuracy, statement coverage, branch coverage, and mutation score~\cite{huang2026ult}. \textsc{TestForge} evaluates an agentic test-generation framework using $\text{pass@1}$, line coverage, and mutation score on \textsc{TestGenEval}~\cite{jain2025testforge}. Wang et al. argue that coverage alone can be a weak indicator of fault-detection ability and use mutation score as a stricter evaluation target for LLM-generated tests~\cite{wang2025mutgen}. Their work treats generated tests as software-engineering artifacts and evaluates their practical quality. In our study, test generation serves a different role: Generated assertions are used as one behavioral window into the LLM’s inferred interpretation of a program.

Recent work suggests that code-tuned LLMs encode signals about generated-code correctness in their internal representations before any external test execution occurs. Approaches such as \textsc{Openia} use intermediate representations from code-specialized LLMs to assess whether generated code is correct~\cite{bui2025openia}, while \textsc{Autoprobe} refines this idea by dynamically selecting informative hidden states and applying them to properties such as compilability, functionality, and security~\cite{vu2025autoprobe}. Similarly, Ribeiro et al. study LLMs’ internal representation of code correctness by contrasting hidden states for correct and incorrect code for the same programming tasks, showing that the extracted representation can help select higher-quality code samples without test execution~\cite{ribeiro2026internalcorrectness}. Other work moves toward more mechanistic diagnostics: Sparse-autoencoder analyses identify activation directions associated with correctness~\cite{tahimic2025sae}, while \textsc{CodeCircuit} traces how information flows through the LLM during code generation and uses this analysis to identify which lines of code contribute to correctness-related predictions~\cite{he2026codecircuit}. Together, these studies show that correctness-relevant information is reflected in internal representations and can be used to analyze, predict, or influence the quality of generated code. Despite this progress, this line of work has largely centered on correctness as the target property, while internal mechanisms for handling other aspects of program comprehension---such as conflicts between multiple sources of information---remain comparatively underexplored.

\subsection{Mechanistic Interpretability of Language Models}
\label{sec:rw-mechanistic}

Mechanistic interpretability aims to explain LLM behavior by analyzing internal components and representations. Causal mediation analysis introduced the idea of treating hidden units, attention heads, or layers as mediators between input and output and intervening on them to estimate causal contribution~\cite{vig2020causal}. Causal abstraction and interchange-intervention frameworks provide a more general account of how neural computations can be related to abstract, human-interpretable causal explanations of LLM behavior~\cite{geiger2025causal}. In transformer language models, causal tracing and activation patching have become standard tools for localizing behaviorally relevant activations. Meng et al. use causal tracing to locate factual associations in GPT models~\cite{meng2022locating}. Wang et al. use activation patching and path patching to identify an indirect-object-identification circuit in GPT-2 Small~\cite{wang2023ioi}. Conmy et al. systematize this workflow through automated circuit discovery~\cite{Conmy2023}.

Activation patching is particularly relevant to our work, as it provides a causal method for asking whether a specific internal activation contributes to an externally observable behavioral difference. Methodological work emphasizes that patching results depend on the choice of source and destination prompts, corruption or contrast construction, patching granularity, and behavioral metrics: Zhang and Nanda show that activation-patching results can vary substantially with metric and method choices~\cite{zhang2024bestpractices}. Heimersheim and Nanda emphasize that patching should be interpreted as evidence of causal contribution under a specified setup, not as a complete explanation of a models’s algorithm~\cite{heimersheim2024patching}. These cautions are especially important when patching code or other non-canonical natural-language tasks.

The residual stream is a natural target for such interventions because work on transformer circuits conceptualizes it as the main communication channel through which layers read, write, and accumulate information across layers~\cite{elhage2021framework}. Residual-stream interventions are complementary to sparse-feature approaches, which seek to decompose dense activations into interpretable features using sparse autoencoders~\cite{gao2025scaling}. In software engineering, sparse-autoencoder and attribution-graph approaches have recently been used to analyze correctness-related representations in code LLMs~\cite{tahimic2025sae, he2026codecircuit}. Our work builds on these mechanistic tools: We use causal residual-stream patching on controlled code inputs to study how LLMs reconcile competing sources of program semantics.

\subsection{Semantic Conflicts in Software Artifacts}
\label{sec:rw-semantic-conflicts}

Software artifacts combine executable code, which determines runtime behavior, with non-executable semantic cues such as comments, docstrings, and identifiers that convey intent or rationale.
Prior work has emphasized the importance of comments and natural-language artifacts for program comprehension and maintenance, while also showing that they can diverge from code---for example, when comments are not updated alongside code changes~\cite{mastropaolo2021commentcompletion,steiner2022commentinconsistency, Abdelsalam2026EffectComments} or when linked resources decay over time~\cite{hata2019links}.
These studies motivate treating executable code and non-executable cues as related but separable sources of information.

The distinction between executable behavior and natural-language intent is also reflected in code-tuned LLM benchmarks.
\textsc{HumanEval} tasks are specified through natural-language doc strings and evaluated by tests~\cite{chen2021evaluating}; \textsc{MBPP} tasks are specified by natural-language descriptions and example tests~\cite{austin2021program}, which mirrors ordinary software-engineering practice, where natural language and code jointly specify intended behavior.
However, it also means that standard end-to-end benchmarks usually evaluate whether an LLM satisfies the provided task specification, not how the LLM internally balances executable evidence against non-executable semantic cues in case of conflict/inconsistency.
This creates an opportunity and a need at the same time for mechanistic analysis: Semantic conflicts in code provide controlled contrasts for studying how LLMs process different sources of program behavior.

\subsection{Positioning of Our Work}
\label{sec:positioning}

Our work sits at the intersection of these three research areas. Unlike prior internal-correctness studies that primarily compare correct and incorrect generations, we study prompts in which the input itself contains conflicting evidence about behavior. Unlike standard code-generation or test-generation evaluations, we do not only ask whether the final artifact is correct, but we ask how information from different parts of the prompt causally contribute to the LLM’s behavior. We therefore use semantic conflicts in code as the experimental subject for mechanistic analysis, and compare two behavioral endpoints: final-output prediction and generated unit-test assertions. This positions our study as a causal analysis of how LLMs route executable and non-executable information when interpreting code, providing a more fine-grained view of how competing signals are resolved during inference.
\section{Research Questions}
\label{sec:research_questions}

Based on our overarching goal, we start from a setting where we have Python code snippets with aligned and conflicting versions at our disposal, allowing us to test whether conflicting semantic cues cause the LLM to deviate from execution-grounded behavior. To provide an overview of how the conflicts affect the LLM in standard program-comprehension tasks, we pose our first research question:

\researchquestion{RQ\textsubscript{1.1}}{How do semantic conflicts affect LLM behavior during program comprehension?}

To answer this question, we create prompts with the snippets and an output prediction task and analyze the output tokens of the LLM compared to the snippet's execution behavior. Based on that knowledge, we aim to next identify token-layer sites that shift the LLM's output preference and pose the second part of the research question:

\researchquestion{RQ\textsubscript{1.2}}{Which token-layer sites causally contribute to shifting LLM outputs between aligned and conflicting inputs?}

For this question, we store the internal representations gained through RQ\textsubscript{1.1} and perform residual activation patching per layer and token, where we patch the information from conflicting code versions into the aligned one and vice versa. We execute the LLM with the patched representation and analyze whether the patch changes the model's output preference.

In addition to the output prediction task, we also analyze the LLM behavior and internal representations for a standard downstream task in software engineering: test generation. Thus, we pose our second research question:

\researchquestion{RQ\textsubscript{2.1}}{How do semantic conflicts affect LLM behavior during test generation?}

To answer this question, we create prompts with the snippets, ask the LLM to generate \textsf{pytest} style unit tests with 3~test cases, and execute them to assess their alignment with semantic cues and implementation. 

\researchquestion{RQ\textsubscript{2.2}}{Which token-layer sites causally contribute to differences in LLM-generated unit tests between aligned and conflicting inputs?}

In line with RQ\textsubscript{1.2}, we create assertion-specific prompts in which each generated test case is provided up to the assertion value. We then perform residual activation patching per layer and token and analyze whether the patch changes the model's completion at the assertion value.

\section{Methodology}
\label{sec:methodology}

To answer our research questions, we create a curated set of Python snippets, prompt the LLM for final-output and unit-test generation, and perform residual activation patching to identify token-layer sites which shift the LLM's output preference.

\subsection{Independent Variables and Code Snippets}

Our goal is to identify how LLMs handle conflicts between non-executable semantic cues, such as function names or comments, and executable program behavior.
In this context, we construct systematically varied Python snippets and categorize them into \textit{aligned} and \textit{conflicting}, depending on whether the non-executable semantic cues and the executable program behavior are consistent with each other.
We consider one independent variable (``cue-implementation alignment'') with 3~levels:
For the \alignedSn{} snippets, the non-executable semantic cues and the executable program behavior imply the same outcome.
For the \cuevaried{} snippets, we introduce a small, localized change to a semantic \textbf{cue} that contradicts the executable program's behavior (e.g., replacing ``even'' with ``odd''; see~\autoref{fig:ExperimentOverview}).
For the \implementationvaried{} snippets, we instead introduce a localized \textbf{code}~change that alters the program's behavior to deviate from the intent suggested by the semantic cue (e.g., replacing ``0'' with ``1''; see~\autoref{fig:ExperimentOverview}).

In total, we created 45 Python snippet triplets (i.e., $135$ total snippets; similar to~\autoref{fig:ExperimentOverview}) consisting of short functions with meaningful identifier names and, where necessary, a function-level comment describing the intended behavior.
Each stimulus pair differs by 1--2 localized changes while ensuring that all regarded LLMs tokenize the changes in the exact same number of tokens, so that the post-change tokens remain aligned between prompts.
To ensure meaningful contrasts, each pair must incite a cue–implementation conflict for, at least, one studied LLM for output generation.
We provide details on the snippets and their construction in the replication package~\cite{zenodo:dataset}.

\subsection{LLMs}

We selected four open-weight autoregressive transformer LLMs that support residual-stream analysis and can follow the task prompts. The selected LLMs are \textsc{CodeLlama-7B-Python}~\cite{Roziere2024Codellama} and the \textsc{Instruct} variants of \textsc{Qwen2.5-7B}~\cite{Qwen2}, \textsc{Mistral-7B}~\cite{Jiang2023Mistral7b}, and \textsc{Llama-3.1-8B}~\cite{Llama3}. This set includes a code-tuned LLM and general-purpose, code-capable LLMs and is available for mechanistic analysis through \textsc{TransformerLens}~\cite{Nanda2022TransformerLens}. \textsc{Qwen2.5} contains $28$ transformer layers, whereas the other LLMs contain $32$.

\subsection{Tasks}

We investigate two complementary programming tasks: final-output prediction and unit-test generation.

\paragraph{Final-Output Prediction (RQ\textsubscript{1})}
\label{sec:method-rq11}

The final-output prediction task evaluates the LLM's direct answer to an established program-comprehension question by predicting the output of a given snippet. We execute the LLM with zero temperature and validate the generated predictions against the executed runtime output to assess the consistency of the LLM with the implementation. We measure cue consistency by comparing against the predefined behavior suggested by the cue or paired aligned variant. Thus, a response can be \bothconsistent{}, \cueconsistent{}, \executionconsistent{}, or neither-consistent. The ground truth for correctness is the execution behavior.

\paragraph{Unit-Test Generation (RQ\textsubscript{2})}
\label{sec:method-rq21}

The unit-test generation task measures the behavioral interpretation that the LLM externalizes as \textsc{pytest}-style unit-test assertions.
This task requires the LLM to formulate expected behavior in executable test form with three test cases for the function \verb|compute|. 
The generated unit tests are treated as a behavioral probe of the LLM's inferred interpretation of the snippet. 
To determine whether an assertion aligns with the semantic cues or the implementation, we distinguish between the two types of conflicts.
For cue-varied prompts, we evaluate the conflicting generated assertions against a cue-semantics oracle (\cueconsistent{}) and the conflicting implementation (\executionconsistent{}).
For implementation-varied prompts, we evaluate the assertions against the conflicting (\executionconsistent{}) and the aligned implementation (\cueconsistent{}).
If a test passes both implementations for a varied prompt, it is both/non-discriminating, as this test does not discriminate between the two interpretations.
For aligned prompts, we evaluate the assertions against the aligned implementation (\bothconsistent{}).
Otherwise, the assertion is labeled as \neither{}.

For RQ\textsubscript{2.2}, we create one prompt for each generated test case, in which we separate the assertion expression from its generated expected value, so that \verb|assert compute(4) == | is part of the prompt and the value after \verb|==| is the output contrast. Based on this prompt, we patch the LLM and analyze how the output preference between the two expected values changes.

\subsection{Statistical Analysis on Behavior (RQ\textsubscript{1.1} and RQ\textsubscript{2.1})}

To assess whether semantic conflicts affect LLM behavior, we use standard paired statistical tests to compare between aligned and conflicting versions of the snippet triplets. The paired design controls for differences in snippet complexity.

For the final-output prediction, the outcome is binary (i.e., correct or incorrect). Thus, we use exact paired McNemar tests, which are specifically designed for paired binary data~\cite{McNemar1947}. We report the paired risk difference as aligned minus conflicting correctness and measure the effect size using Cohen's g~\cite{Cohen1977}. Because the comparisons are paired by snippet, they control for differences in baseline snippet difficulty and isolate the effect of introducing a semantic conflict.

For unit-test generation, we evaluate LLM behavior using the pass rate of the 3 assertions per snippet. Since these rates are not binary and may not be normally distributed, we use paired Wilcoxon signed-rank tests~\cite{Wilcoxon1945} and measure the effect using signed rank-biserial effect size $r_{\mathrm{rb}}$.

Finally, since our analyses include multiple LLMs and conflicts, we apply a Holm correction to control for family-wise error~\cite{Holm1979}, and use the typical $\alpha=0.05$ threshold. This holds for all statistical analyses, including RQ\textsubscript{1.2} and RQ\textsubscript{2.2}.

\subsection{Residual Stream Activation Patching (RQ\textsubscript{1.2} and RQ\textsubscript{2.2})}
\label{sec:MethodActivationPatching}

We use residual stream activation patching to identify token-layer sites that causally shift LLM behavior between our paired snippets.
For each snippet pair, the LLM is run once on each snippet and the residual stream activations are cached per patch site. A patch site is defined by a transformer layer and an aligned patch unit (e.g., \verb|even|, \verb|odd|, \verb|0|, \verb|1|, see~\autoref{fig:ExperimentOverview}). Consecutive changed tokens are treated as one combined patch unit to preserve the atomicity of the manipulation.

Then, we execute patched runs by replacing the residual stream activation of a patch site in a destination prompt with the corresponding activation from the paired source prompt ($\text{source} \rightarrow \text{dest}$). 
We perform patching starting from the first token position where the snippets differ, to ensure that patched activations occur at positions that can causally depend on the manipulation.
We patch in both directions independently, fixing the conflicting prompt (forward: $\text{aligned}\rightarrow\text{conflicting}$) and corrupting the aligned prompt (reverse: $\text{conflicting}\rightarrow\text{aligned}$). 

For each patching comparison consisting of snippet pair and task, we define a \textit{source-associated behavior} and a \textit{destination-associated behavior} from the two semantic outputs represented by the stimulus pair.
We interpret a patched activation as causally contributing to the paired contrast when it shifts the LLM's output preference toward the source reference. This contrast is defined independently of the LLM's generated response label.

For RQ\textsubscript{1.2}, we contrast the two final-output candidates associated with a snippet pair. 
For RQ\textsubscript{2.2}, patching identifies prompt and generated-prefix states that shift the expected value of a generated assertion. We use the expected values used for labeling the assertions in RQ\textsubscript{2.1} now to contrast between aligning and conflicting snippets in a pair. Assertions for which both candidates are identical are excluded in RQ\textsubscript{2.2} because they do not distinguish the conflict.

\subsubsection{Evaluation of LLM Patching}

In line with prior work~\cite{zhang2024bestpractices, wang2023ioi}, we compute how much a patched run changes from original destination behavior to source behavior using a recovery score. This score compares the logits of candidate outputs associated with each prompt compared to the unpatched baseline logits. A larger recovery score indicates that the patched residual state activation has a stronger causal influence on the LLM's output, identifying contributing token-layer sites.
To focus on patch sites with a considerable causal effect, we exclude all patch sites with a recovery below $0.3$. 
We report additional results with thresholds of $0.2$ and $0.5$ as sensitivity checks in the replication package~\cite{zenodo:dataset}. 

The denominator of the recovery score is the unpatched source--destination margin gap, $\Delta=m_S-m_D$. If this denominator is close to zero, normalized recovery can look large even when the raw logit change caused by the patch is modest. We therefore perform a robustness check that filters runs with small denominators by requiring $|\Delta|\geq\tau$ for $\tau\in\{0.05, 0.10, 0.25\}$ logits, with $\tau=0.10$ as the primary threshold. The exact formulas and edge-case handling are in the replication package~\cite{zenodo:dataset}.

Next, we localize the patch sites with the highest causal impact.
One aspect we analyze is how causal contributions are distributed across different parts of the input.
In addition to changed region and readout sites, we identify all intermediate patch units exerting a substantial causal effect, which we call \emph{carrier tokens}. For RQ\textsubscript{2.2}, we additionally differentiate between carrier tokens within the prompt (prompt carriers) and within the generated test case (response carriers).
To identify the best recovery per such category of patch units, we take the maximum recovery over layers.
We analyze the strength of recovery over categories of patch units for every model, conflict type and patch direction using the paired Wilcoxon signed-rank tests~\cite{Wilcoxon1945} test and describe the signed rank-biserial effect size $r_{\mathrm{rb}}$.

Furthermore, we calculate the best recovery layer for every category of patch units. 
We take the unit with the largest recovery score and record the layer at which it peaks.
To account for the varying number of transformer layers (Qwen2.5~7B:~28; other LLMs:~32), we normalize the layer indices to their relative depth in the interval of $[0,1]$.
We analyze the best recovery layer over categories of patch units for every model, conflict type and patch direction using the paired Wilcoxon signed-rank tests~\cite{Wilcoxon1945} test and describe the signed rank-biserial effect size $r_{\mathrm{rb}}$.

\section{Results}
\label{sec:results}

In this section, we present the results of our behavioral and causal analyses structured along our two tasks.

\subsection{Effect of Semantic Conflicts on LLM Behavior (RQ\textsubscript{1.1})}

Final-output prediction was the first behavioral test of the constructed semantic conflicts. We compare aligned and conflicting prompts using execution-grounded correctness. Our results show that correctness is consistently higher for aligned prompts than for conflicting prompts across all LLMs, with an average drop of $39.7$ percentage points. When conducting paired McNemar tests, every LLM--conflict comparison remains statistically significant after Holm correction (see Table~\ref{tab:rq11-model-stats}), with effect sizes between $0.3$ and $0.5$.

\begin{table}[t]
  \centering
  \caption{RQ\textsubscript{1.1} Paired exact McNemar tests comparing aligned and conflicting final-output correctness.}
  \resizebox{\linewidth}{!}{%
  \begin{tabular}{llccccr}
    \toprule
    LLM & Conflict Family & \multicolumn{3}{c}{Correctness} & Effect Size & $p_{\mathrm{Holm}}$ \\
    & & Aligned & Conflicting & Diff. & $g$ &  \\
    \midrule
    \multirow{2}{*}{CodeLlama 7B} & Cue-varied & \multirow{2}{*}{88.9\%} & 44.4\% & 44.4 & 0.50 & $<0.001$ \\
     & Impl-varied &  & 42.2\% & 46.7 & 0.46 & $<0.001$ \\
    \addlinespace[.25em]
    \multirow{2}{*}{Llama 3.1 8B} & Cue-varied & \multirow{2}{*}{84.4\%} & 64.4\% & 20.0 & 0.50 & \textbf{$0.004$} \\
     & Impl-varied &  & 44.4\% & 40.0 & 0.41 & $<0.001$ \\
    \addlinespace[.25em]
    \multirow{2}{*}{Mistral 7B} & Cue-varied & \multirow{2}{*}{73.3\%} & 33.3\% & 40.0 & 0.45 & \textbf{$<0.001$} \\
     & Impl-varied &  & 33.3\% & 40.0 & 0.30 & \textbf{$0.003$} \\
    \addlinespace[.25em]
    \multirow{2}{*}{Qwen2.5 7B} & Cue-varied & \multirow{2}{*}{77.8\%} & 35.6\% & 42.2 & 0.41 & \textbf{$<0.001$} \\
     & Impl-varied &  & 33.3\% & 44.4 & 0.46 & \textbf{$<0.001$} \\
    \bottomrule
  \end{tabular}}
  \label{tab:rq11-model-stats}
\end{table}

\begin{figure}[t]
  \centering
  \includegraphics[width=\linewidth]{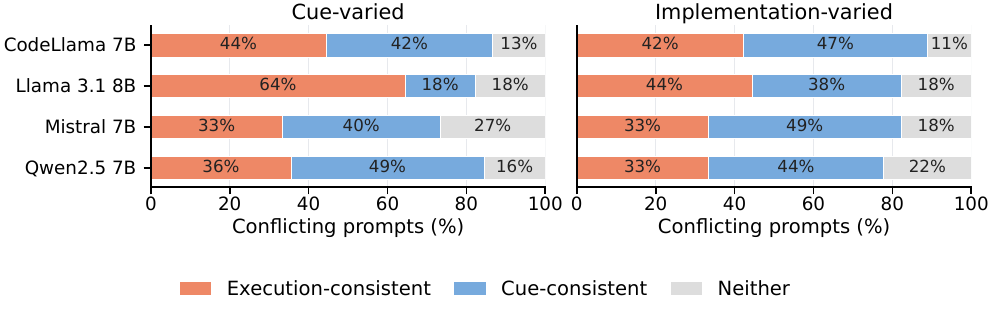}
  \caption{RQ\textsubscript{1.1} Distribution of response labels on conflicting prompts. Cue-consistent responses are incorrect under the execution-grounded evaluation but indicate sensitivity to the misleading cue.}
  \label{fig:rq11-labels}
\end{figure}

To characterize the errors induced by the conflicts, we further divide the incorrect conflicting-prompt responses based on whether they match the misleading cue to separate conflict-induced failure (\cueconsistent{} output) from generic task failure (\neither{}). As shown in Figure~\ref{fig:rq11-labels}, the division shows that many erroneous outputs of the LLMs are \cueconsistent{} (up to~$49\%$), even frequently surpassing the rate of correct \executionconsistent{} outputs (above $33\%$). Thus, the LLMs exhibit a directed bias under conflict, with many failures aligning with the incorrect but semantically salient cue.

\RQAnswer{RQ\textsubscript{1.1}}{Conflicting semantic cues reliably reduce final-output correctness. The effect is large, statistically significant for every LLM--conflict pair, and frequently directed toward the misleading cue.}

These findings establish that the dataset contains effective conflicts motivating the residual-stream analysis in RQ\textsubscript{1.2}.

\subsection{Internal Localization of Semantic-Conflict Information During Output Prediction (RQ\textsubscript{1.2})}

\begin{figure}[t]
  \centering
  \includegraphics[width=\linewidth]{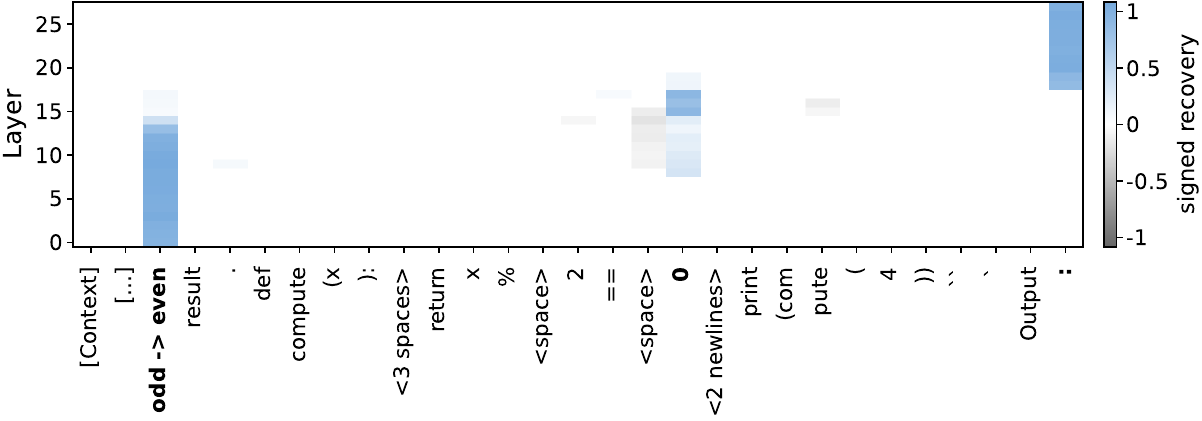}
  \caption{RQ\textsubscript{1.2} Representative residual-recovery heatmap for Qwen2.5 7B, cue-varied pair $001$, patching the conflicting-cue source into the aligned-cue destination. The changed cue token recovers in early layers, token \texttt{0} exhibits a later recovery band, and the readout token recovers in the final layers.}
  \label{fig:rq12-example-heatmap}
\end{figure}

Having established that semantic conflicts affect final-output behavior, we investigate in RQ\textsubscript{1.2} where the conflicting information becomes causally available inside the LLM. 
To illustrate the results, we present a representative run for Qwen2.5 7B on cue-varied pair $001$ in Figure~\ref{fig:rq12-example-heatmap}. The execution-grounded result is \textsf{True}, whereas the conflicting cue implies \textsf{False}. The strongest recovery score first appears directly at the changed cue token \textsf{odd} in an early layer. A second region with a high recovery score appears at the token \textsf{0}, which carries part of the relevant parity predicate. A high recovery score finally appears at the last readout token, where the output is decoded. 

\begin{figure}[t]
  \centering
  \includegraphics[width=\linewidth]{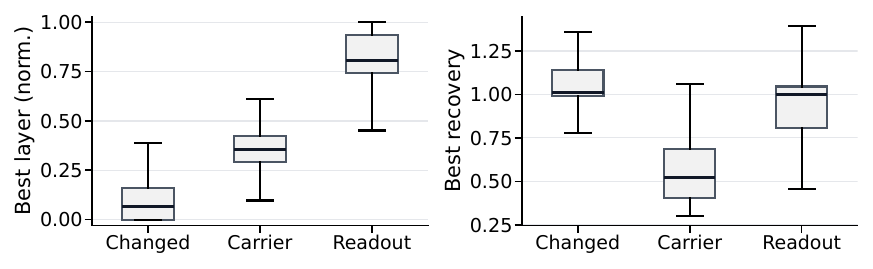}
  \caption{RQ\textsubscript{1.2} Site-group comparison at the primary carrier threshold. Changed region peaks early, carrier tokens in middle layers, and readout site late; recovery is strongest at changed and readout sites and substantial for carriers.}
  \label{fig:rq12-site-groups}
\end{figure}

This example is representative of the mechanistic pattern we identified over all patch runs: There is a noticeable recovery at the changed patch site, at a few intermediate code tokens and at the last token (i.e., the readout site).
Our intermediate token analysis identified 2\,640 carrier tokens across 720 runs, with a mean of 3.67 and a median of one carrier token per run.
This sparsity indicates that conflict-relevant residual information is not uniformly available at all downstream tokens. Instead, it concentrates in a small number of intermediate tokens before being aggregated at the readout site. 
Analyzing the recovery distribution across layers shows the same staged pattern. Median best recovery occurs around layer 2 at changed-region sites, around layers 10--11 at carrier tokens, and around layers 23--26 at readout sites (see Figure~\ref{fig:rq12-site-groups}). Thus, carrier tokens peak several layers after the changed region but well before the late readout stage near generation. These layer offsets are pairwise significant with extremely large effects ($p_{\mathrm{Holm}}<0.001$, $r_{\mathrm{rb}}\geq0.958$). A detailed overview is provided in the replication package~\cite{zenodo:dataset}.

Analyzing the recovery strength shows that carrier tokens exert a substantial effect (median $0.52$), but it is lower than for the changed-region (median $1.01$) and readout site (median $1.00$), as shown in Figure~\ref{fig:rq12-site-groups}. Pairwise tests show that there is a significant difference between all three categories ($p_{\mathrm{Holm}}<0.001$), with a large recovery-strength contrast between carrier token and both changed-region and readout site ($r_{\mathrm{rb}}\geq0.892$, $r_{\mathrm{rb}}\geq0.830$). The effect size between changed-region and readout site is much smaller (median difference $<0.1$, $r_{\mathrm{rb}}\geq0.371$).
Recovery strength, carrier prevalence, and staging remain similar across both carrier-threshold sensitivity checks and forward/reverse patching comparisons. Detailed analyses are included in the replication package~\cite{zenodo:dataset}.

\RQAnswer{RQ\textsubscript{1.2}}{Residual patching provides a mechanistic account of the conflict in RQ\textsubscript{1.1} and reveals a general causal staging pattern for semantic conflicts: The changed cue/code region have a high recovery in early layers, sparse carrier tokens at intermediate layers, and the readout site in layers near generation. Thus, conflict information is introduced at the edited semantic site, made available at selected intermediate program tokens, and accumulated at the generation context.}

\subsection{Effect of Semantic Conflicts on Downstream Tasks (RQ\textsubscript{2.1})}

Next, we study the effect of semantic conflicts on generating unit tests as a downstream behavioral task common in software engineering. This task is more open-ended than final-output prediction because an LLM can generate different test inputs and multiple assertions per snippet. Nevertheless, the behavioral effect observed in RQ\textsubscript{1.1} also appears for unit-test generation, with lower assertion pass rate under both conflict pairs for every LLM.

All eight LLM--conflict comparisons show statistically significant reductions in assertion pass rate after Holm correction as shown in Table~\ref{tab:rq21-model-stats}. Aligned assertions result in a pass rate between $77\%$ to almost $90\%$. In comparison, the cue-varied pass rate is reduced by $18.5$ to $31.9$ percentage points; this reduction is generally smaller than seen in RQ\textsubscript{1.1}. In contrast, implementation-varied reductions remain similarly large, ranging from $30.3$ to $43.9$ percentage points. The effects are large to very large ($r_{\mathrm{rb}}=0.65$--$1.00$). The conflicts therefore affect not only direct output extraction but also the behavioral specifications encoded in generated tests.

\begin{table}[t]
  \centering
  \caption{RQ\textsubscript{2.1} Paired Wilcoxon tests comparing aligned and conflicting generated-test assertion correctness.}
  \resizebox{\linewidth}{!}{%
  \begin{tabular}{llrrrcr}
    \toprule
    LLM & Conflict Family & \multicolumn{3}{c}{Assertion Pass Rate} & Effect Size & $p_{\mathrm{Holm}}$  \\
    & & Aligned & Conflicting & Diff. & $r_{\mathrm{rb}}$ & \\
    \midrule
    \multirow{2}{*}{CodeLlama 7B} & Cue-varied & 81.5\% & 49.6\% & 31.9 & 0.82 & $<0.001$ \\
     & Impl-varied & 81.1\% & 37.1\% & 43.9 & 0.89 & $<0.001$ \\
    \addlinespace[.25em]
    \multirow{2}{*}{Llama 3.1 8B} & Cue-varied & 89.6\% & 71.1\% & 18.5 & 1.00 & $<0.001$ \\
     & Impl-varied & 89.6\% & 48.5\% & 41.1 & 1.00 & $<0.001$ \\
    \addlinespace[.25em]
    \multirow{2}{*}{Mistral 7B} & Cue-varied & 77.8\% & 50.7\% & 27.0 & 0.81 & $<0.001$ \\
     & Impl-varied & 77.8\% & 34.1\% & 43.7 & 0.91 & $<0.001$ \\
    \addlinespace[.25em]
    \multirow{2}{*}{Qwen2.5 7B} & Cue-varied & 83.0\% & 62.5\% & 20.5 & 0.65 & $0.002$ \\
     & Impl-varied & 83.0\% & 52.7\% & 30.3 & 0.71 & $<0.001$ \\
    \bottomrule
  \end{tabular}}
  \label{tab:rq21-model-stats}
\end{table}

\begin{figure}[t]
  \centering
  \includegraphics[width=\linewidth]{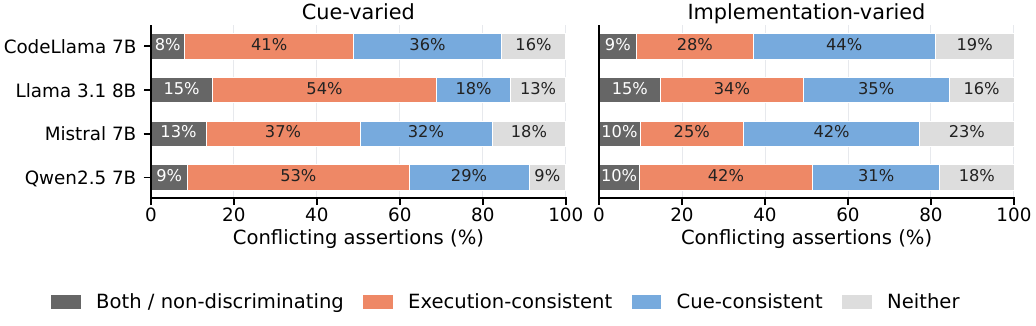}
  \caption{RQ\textsubscript{2.1} Assertion labels for conflicting generated unit tests, in which both/non-discriminating assertions are runtime-correct assertions and share the same expected value under both execution semantics and misleading cue.}
  \label{fig:rq21-labels}
\end{figure}

The generated unit tests show that, across both conflicts, many assertion failures are \cueconsistent{}~($18\%$--$42\%$) rather than arbitrary ($9\%$--$23\%$; see Figure~\ref{fig:rq21-labels}).
Furthermore, a considerable portion of assertions are both/non-discriminating, which means the selected values fail to expose the underlying conflict (up to $15\%$).
This mirrors the pattern of RQ\textsubscript{1.1} at the level of generated behavioral specifications, and additionally shows that some generated tests avoid the conflict by choosing non-diagnostic inputs, which is highly problematic in practice (see Section~\ref{sec:discussion-synthesis})

\RQAnswer{RQ\textsubscript{2.1}}{Semantic conflicts also reduce the pass rate for assertions during unit-test generation. The effect is significant across all LLM--conflict pairs, and many erroneous tests follow the conflicting cue. A separate both/non-discriminating category shows that some runtime-correct generated tests choose inputs that avoid the conflict.}

\subsection{Localization of Semantic-Conflict Information During Unit-Test Generation (RQ\textsubscript{2.2})}

In RQ\textsubscript{2.2}, we investigate how semantic conflicts affect the expected assertion-value that the LLM writes into a unit test.

\begin{figure}[t]
  \centering
  \includegraphics[width=\linewidth]{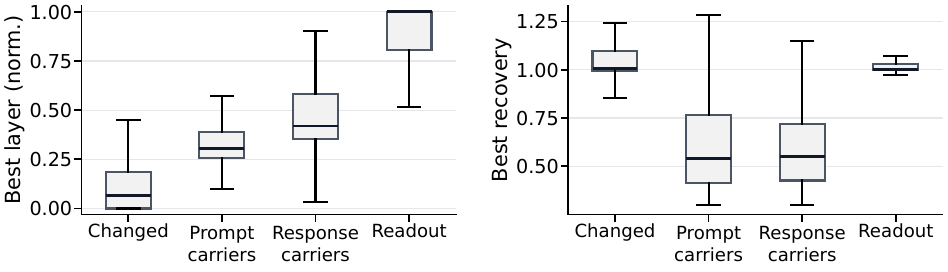}
  \caption{RQ\textsubscript{2.2} Assertion-specific staging and recovery strength of conflict information during unit-test generation at the primary carrier threshold.}
  \label{fig:rq22-site-groups}
\end{figure}

Analyzing the recovery distribution across layers shows a staged pattern that closely parallels final-output prediction while adding a distinct response-generation stage before the readout site. Median best recovery occurs around layer 2 at changed-prompt sites, around layers 8--10 at prompt carriers, around layers 11--13 at response carriers, and at the final layer (27 or 31) at the expected-value readout (Figure~\ref{fig:rq22-site-groups}). All adjacent layer offsets are significant with large to very large rank-biserial effects: prompt carriers follow changed-prompt sites by a median normalized-depth offset of $0.222$ ($p_{\mathrm{Holm}}<0.001$, $r_{\mathrm{rb}}=0.952$), response carriers follow prompt carriers by $0.111$ ($p_{\mathrm{Holm}}<0.001$, $r_{\mathrm{rb}}=0.676$), and the readout follows response carriers by $0.419$ ($p_{\mathrm{Holm}}<0.001$, $r_{\mathrm{rb}}=0.982$).

Carrier prevalence remains sparse, with prompt carriers averaging 2.35 tokens or 5.3\% of eligible patch units, and response carriers averaging 1.64 tokens or 4.6\% of eligible patch units.
The share between prompt and response carriers does not differ significantly ($p_{\mathrm{Holm}}=0.475$), indicating that the smaller response count mainly reflects fewer available positions.

Recovery strength differs in line with the recovery-strength pattern observed in RQ\textsubscript{1.2}: Median best recovery is high for both the changed region and expected-value readout ($1.01$ and $1.00$), but lower for both carrier sites ($0.54$).
Pairwise statistical tests confirm very large recovery-strength contrasts between carrier and changed-prompt respective expected-value readout sites ($p_{\mathrm{Holm}}<0.001$, $r_{\mathrm{rb}}\leq-0.919$, respective $p_{\mathrm{Holm}}<0.001$, $r_{\mathrm{rb}}\geq0.876$). In contrast, prompt and response carriers do not differ significantly ($p_{\mathrm{Holm}}=0.913$), as well as changed region and readout site ($p_{\mathrm{Holm}}=0.355$).
Forward and reverse patching preserve the same four-part staging, with small direction-specific offsets reported in the replication package~\cite{zenodo:dataset}.

\RQAnswer{RQ\textsubscript{2.2}}{Unit-test generation exhibits a four-stage localization pattern similar to what we observed for output prediction in RQ\textsubscript{1.2}. Conflicting information is first recoverable at the changed prompt, then at sparse prompt carriers, then at generated assertion-prefix carriers, and finally at the expected-value readout. Recovery strength is highest at the changed prompt and readout, while prompt and response carriers show lower but comparable recovery.}

\section{Discussion and Implications}
\label{sec:discussion}

In this section, we discuss our results and their implications.

\subsection{Synthesis and Interpretation}
\label{sec:discussion-synthesis}

Our results in RQ\textsubscript{1.1} show that semantic conflicts expose a specific failure mode in which LLMs frequently follow cue-consistent behavior when implementation and semantic cues disagree. This decision between cue- and execution-consistent behavior occurs in both output prediction and unit-test generation, propagating conflicts into a downstream artifact that can later misguide developers, tools, or further LLM actions.

Beyond cue-consistency failures, our results in RQ\textsubscript{2.1} highlight another problematic bias: Many non-discriminating unit tests illustrate how conflicts can persist unnoticed, since these tests avoid the semantic disagreement.
Unlike a failing test, a passing but non-discriminating test may not alert the developer that the implementation diverges from the intended behavior, leaving the underlying disagreement hidden and unresolved.

These observations have implications for downstream software-engineering tasks that rely on LLMs inferring developer intent correctly, including code review, automated repair, refactoring, test generation, and agentic workflows.
Following the wrong source of information in a conflict can produce outputs that are plausible and useful-looking while preserving or introducing the wrong behavior. Our findings therefore provide empirical evidence from a new angle for the usual maintainability argument for accurate documentation and naming~\cite{Aghajani2020}. Bad comments, stale documentation, and misleading identifiers are not merely unhelpful context for an LLM, but they can actively steer generated code, reviews, repairs, or tests toward an incorrect interpretation. Semantic-cue sensitivity is therefore both useful and hazardous. LLMs need such cues to infer intent, but AI-assisted workflows need checks that identify when those cues conflict with execution.

\subsection{Mechanistic Interpretability and Causal Analysis}
\label{sec:discussion-mechanistic-interpretation}

For RQ\textsubscript{1.2} and RQ\textsubscript{2.2}, we turn the behavioral effects of LLMs into causal study objects using a mechanistic analysis through interventions.  Although mechanistic interpretability methods are well established for natural-language behavior~\cite{meng2022locating,wang2023ioi}, they have rarely been applied in software engineering~\cite{ribeiro2026internalcorrectness, he2026codecircuit}. Our experimental framework demonstrates how these methods can be adapted to problems in software engineering with the example of semantic conflicts in program code. 

Our study demonstrates how residual-stream activation patching serves as a targeted discovery mechanism that revealed a staged localization pattern across the prompt and generation context, which gives the behavioral result a more actionable form: It identified at which layers and tokens the conflict-relevant information is recoverable and where follow-up methods should look for the components that carry cue or execution information, thereby reducing the otherwise large search space to a targeted set of candidate token-layer sites.

This way, our framework enables more targeted follow-up analyses.
For example, path patching can trace whether information from the changed region reaches intermediate tokens and readout sites through specific components, while component-level ablations can test whether those components are necessary for cue- or execution-following behavior. Validated sites can further support cross-model comparisons and monitoring of conflict-related internal features during code tasks.

\subsection{Implications for Research}
\label{sec:discussion-researchers}

For researchers, the main implication is methodological. Activation patching provides a way to connect behavioral evaluation with causal evidence about where task-relevant information is recoverable inside the LLM. Our results show that this approach can be applied to software-engineering problems when token-aligned inputs, well-defined behavioral contrasts, and localized changes in the prompt are available.

Semantic conflicts should therefore be understood as a testbed rather than an endpoint. Similar paired designs could study localized code phenomena, such as control-flow conditions, loop bounds, non-local dependencies, API choices, or exception handling behavior. The same methodology could also be applied to other software-engineering tasks and artifacts (design, requirements, etc.) when the task can be framed around a controlled, scoreable contrast. In each case, the contrast must define the output alternatives that are being compared so that behavioral changes and activation-patching effects can be measured.

More broadly, software-engineering research can benefit from adapting mechanistic interpretability methods, which are well established in natural-language settings, ranging from activation probes and sparse autoencoders to activation patching, path patching, targeted ablations, and circuit-discovery techniques~\cite{zhang2024bestpractices, heimersheim2024patching, Conmy2023}. 
Our experimental framework illustrates the feasibility of this transfer and highlights the non-trivial methodological requirements.
Code tasks often require preserving syntactic validity, aligning interventions with meaningful program locations, and defining behavioral contrasts that are both controlled and semantically interpretable. The snippet-based paired design used here addresses these requirements for semantic conflicts, and illustrates how software-engineering datasets may need to be structured to support causal mechanistic analysis. Broadening such task formulations would make it possible to study how LLMs represent program structure, execution behavior, developer intent, and task-specific decisions with the same causal precision that is increasingly common in natural-language analyses.

\subsection{Implications for Practice}
\label{sec:discussion-practitioners}

For practitioners, the immediate implication is that stale comments, misleading names, and inconsistent docstrings represent risks in AI-assisted development. Beyond affecting human understanding, they can steer LLM-generated code, reviews, repairs, and tests. Cleaning up documentation and naming therefore lowers the chance that an LLM infers the wrong intent from otherwise plausible context.

The broader opportunity is to turn this kind of analysis into generation-time tooling if reliable components or circuits can be identified. Residual-stream patching by itself is a discovery method, but validated circuit-level signals could be monitored during generation. In the semantic-conflict setting, a development environment could then flag reliance on misleading cues or detect generated tests that do not exercise the relevant behavioral distinction. More generally, similar diagnostics could be designed for specific code phenomena, such as side effects, API choices, exception paths, or state updates. When a generation depends on one of these fragile decisions, a tool could trigger targeted execution checks, request additional test cases, or ask the LLM to justify the behavior it assumed.

The same analyses are also relevant for LLM providers. Controlled contrasts can become regression suites for code-capable LLMs, especially when new versions are trained or fine-tuned. They can reveal whether an LLM over-relies on misleading semantic cues, whether execution-grounded behavior improves after fine-tuning, and whether improvements on conflicting cases preserve behavior on aligned cases. More broadly, mechanistic signals could inform data curation, fine-tuning objectives, and deployment audits for LLMs used in software-engineering workflows.

\section{Threats to Validity}
\label{sec:threats}

In this section, we discuss potential threats to validity and how we mitigated them.

\subsection{Construct Validity}

A key threat concerns how we interpret LLM outputs when semantic cues and execution behavior diverge. For output prediction, execution behavior provides a reference for correctness, but binary correctness labels would obscure whether an LLM was influenced by semantic cues. For unit-test generation, no independent specification defines the intended behavior, which requires the LLM to infer it from the code, including comments, identifiers, and implementation details.
We mitigate this threat by distinguishing cue-consistent, execution-consistent, both-consistent, and neither-consistent outputs rather than forcing all outputs into a binary correctness judgment. 

A second threat concerns the granularity of our mechanistic analysis. We use residual-stream activation patching, which identifies token-layer sites that causally affect the LLM's preference for specific outputs. This analysis localizes where these effects arise in the residual stream, but it abstracts away from the specific attention heads, multi-layer-perceptron components, and multi-layer paths that produce or propagate them. Future work shall investigate the feasibility of using path patching~\cite{Goldowsky2023} or circuit-discovery methods~\cite{Conmy2023} to analyze the responsible components and propagation paths in more detail.

\subsection{Internal Validity}

We designed the snippets to isolate conflicts between executable behavior and non-executable semantic cues. Each snippet belongs to a paired triplet with localized, token-aligned changes, which controls for length, structure, and baseline difficulty. We preferred reducing such confounds over maximizing snippet diversity, so the results primarily support controlled semantic-conflict claims.

During dataset construction, each snippet pair had to show the intended behavioral contrast for at least one studied LLM. The same LLM had to answer the aligned variant correctly and produce a cue-consistent error on the conflicting variant, ensuring a meaningful patching contrast.
We do not require this behavior from every LLM on every pair and instead test whether the resulting patterns are stable across LLMs.

For activation patching, we use predefined execution- and cue-consistent candidates, so neither-consistent generations are still analyzed when they define a valid contrast. To avoid inflated normalized recovery when source and destination margins are nearly identical, we require a minimum margin gap of $0.1$ logits and report robustness checks with alternative margin thresholds in the replication package~\cite{zenodo:dataset}. We also patch in both directions, so the reported localization patterns do not rely on a single patching direction.

For generation, we use deterministic decoding ($\mathit{temp}=0$) to remove sampling variance and improve reproducibility~\cite{Ouyang2025}. For unit-test generation, we ask for three assertions to give the model repeated opportunities to express its inferred behavior for assertion-specific patching. Since the model chooses the inputs, aligned and conflicting prompts may yield different assertions. We therefore score each assertion against its generated expected value, and exclude non-contrastive assertions from patching when both interpretations give the same value.

\subsection{External Validity}
\label{sec:external_validity}

Our study provides controlled evidence that semantic conflicts can affect LLM behavior and internal representations. However, the exact error rates, effect sizes, and localization patterns may differ in broader software-engineering settings.

We study 45 Python snippet triplets with clear, localized conflicts. This makes execution- and cue-consistent outcomes unambiguous and keeps execution labels, token alignment, and residual patching feasible.
This design does not cover larger files, repositories, dependencies, complex APIs, side effects, or conflicts spread across multiple functions. Such settings often lack clean contrastive pairs and token alignments, so our causal patching design may not transfer directly.

Real conflicts may also be subtler than our constructed examples, such as outdated documentation for edge cases, identifiers that suggest only part of the behavior, comments that omit side effects, or multiple semantic cues that disagree with each other. We therefore interpret the results as evidence that semantic conflicts can affect LLMs, not that all conflict types have the same magnitude or localization.

Another threat concerns our choice of tasks. We use the commonly used output prediction for verifying understanding of program behavior~\cite{Wyrich2023}. Our setting provides a controlled way to measure whether an LLM follows execution behavior or semantic cues. This setup enables precise and systematic analysis, but it may not reflect how semantic conflicts affect more complex software-engineering tasks. 
To increase practical relevance, we study unit-test generation as a downstream task. This task requires the LLM to express its interpretation of the intended behavior and can be evaluated objectively with respect to executable behavior and cue-suggested behavior.
Our results may not generalize to other downstream tasks such as summarization, bug detection, refactoring, or agentic tool use, where outputs are often longer, evaluation is less objective, and interaction, tools, or execution feedback may change how models rely on semantic cues.

Finally, we analyze four open-weight LLMs around 7--8\,B parameters because residual-stream patching requires mechanistic access and feasible compute. This avoids relying on a single model, but larger, closed, tool-augmented, or differently fine-tuned LLMs may behave differently. Future work shall test whether the effects transfer across model scales, model families, and different training data or fine-tuning choices.

\section{Conclusion}
\label{sec:conclusion}

In this paper, we presented an experimental framework that enabled us to investigate LLM behavior on semantic conflicts, where semantic cues suggest program behavior that differs from the code.
To this end, we performed a controlled, mechanistic study with 45 Python code-snippet triplets across four LLMs to analyze the impact of semantic conflicts on LLM behavior for final-output prediction and unit-test generation. 
Furthermore, we causally located conflict-related information within the LLM representation using residual-stream activation patching. 
Our results show that semantic conflicts bias LLM behavior toward following semantic cues, resulting in incorrect output predictions and generated unit tests that either encode the wrong behavior or avoid the conflict altogether.
Using activation patching, we identified a multi-stage localization pattern for semantic conflicts across output prediction and unit-test generation tasks. Conflict information is recoverable at the changed cue/code region in early layers, at sparse intermediate carrier tokens in both the prompt and generated assertion prefix, and finally at the readout site in late layers near generation.

These findings demonstrate how our experimental framework allows us to not only observe LLM output behavior but also localize information that shifts LLM outputs. 
In particular, the results highlight that under semantic conflicts, LLMs may internalize a specific interpretation of the code and carry it into downstream tasks, leading to incorrect output predictions and generated unit tests that encode the wrong behavior or fail to exercise the conflict. 
This has substantial implications for the reliability of AI-assisted development, where such behavior can make semantic conflicts harder to notice and diagnose.

More broadly, a major contribution of our mechanistic experimental framework is that it can be used to study further software engineering phenomena beyond semantic conflicts. Our setup paves the way for future approaches that detect conflict-relevant internal states before generation.
Future work shall build up on our setup and findings by adopting more targeted mechanistic methods, such as identifying circuits responsible for propagating conflict information.
This provides a starting point for approaches to detect, explain, and mitigate errors in AI-assisted workflows.

\section*{Data Availability}
\label{sec:dataavailability}

Following open science principles~\cite{Mendez:2020:Open}, we openly disclose all snippet triplets, their construction details, raw data, and additional results (e.g., of our sensitivity checks)~\cite{zenodo:dataset}.

\section*{Acknowledgment}

This work has been supported by the European Union as part of ERC Advanced Grant ``Brains On Code'' (101052182).

\bibliographystyle{IEEEtran}
\bibliography{bibliography}

\end{document}